\documentclass[reprint,twocolumn,floats,english,jcp,aip,preprintnumbers,amsmath,amssymb,array,superscriptaddress]{revtex4-1}
\usepackage[T1]{fontenc}
\usepackage[latin9]{inputenc}
\usepackage{verbatim}
\usepackage{float}
\usepackage{graphicx}
\usepackage{color}
\makeatletter
\@ifundefined{textcolor}{}
{%
 \definecolor{BLACK}{gray}{0}
 \definecolor{WHITE}{gray}{1}
 \definecolor{RED}{rgb}{1,0,0}
 \definecolor{GREEN}{rgb}{0,1,0}
 \definecolor{BLUE}{rgb}{0,0,1}
 \definecolor{CYAN}{cmyk}{1,0,0,0}
 \definecolor{MAGENTA}{cmyk}{0,1,0,0}
 \definecolor{YELLOW}{cmyk}{0,0,1,0}
 }

\usepackage{dcolumn}
\usepackage{bm}
\usepackage{enumerate}

\makeatother

\usepackage{babel}

\begin{document}

\title{Theory and simulations of quantum glass forming liquids}

\author{Thomas E. Markland}
\affiliation{Department of Chemistry, Stanford University, Stanford, CA
  94305-5080, United States}

\author{Joseph A. Morrone}
\affiliation{Department of Chemistry, Columbia University, 3000 Broadway,
  New York, New York, 10027, United States}

\author{Kunimasa Miyazaki}
\affiliation{Institute of Physics, University of Tsukuba, Tennodai 1-1-1,
  Tsukuba, 305-8571, Japan}

\author{Bruce J. Berne}
\affiliation{Department of Chemistry, Columbia University, 3000 Broadway,
  New York, New York, 10027, United States}

\author{David R. Reichman}
\affiliation{Department of Chemistry, Columbia University, 3000 Broadway,
  New York, New York, 10027, United States}

\author{Eran Rabani}
\affiliation{School of Chemistry, The Sackler Faculty of Exact Sciences,
  Tel Aviv University, Tel Aviv 69978, Israel}

\date{\today}
\begin{abstract}
A comprehensive microscopic dynamical theory is presented for the
description of quantum fluids as they transform into glasses.  The
theory is based on a quantum extension of mode-coupling theory.  Novel
effects are predicted, such as reentrant behavior of dynamical
relaxation times. These predictions are supported by path integral
ring polymer molecular dynamics simulations.  The simulations provide
detailed insight into the factors that govern slow dynamics in glassy
quantum fluids.  Connection to other recent work on both quantum
glasses as well as quantum optimization problems is presented.
\end{abstract}
\maketitle

\section{Introduction}
\label{sec:introduction}
Understanding the fundamental causes of the dramatic slowdown of
dynamics when a liquid transforms into a glass is still a subject of
great debate.\cite{Stillinger01,Berthier05,Biroli09,Chandler09}
Essentially all discussion of the glass transition has focused on the
strictly classical regime of liquid state behavior, namely where the
de Broglie wave length is significantly smaller than the particle
size.  Given that nearly all known glass forming liquids fall well
within this regime,\cite{Angell96} it is clear that the classical
approximation is generally justified. However there are several
interesting and important examples where quantum fluctuations and
glassiness coexist.\cite{Wu91,Imry09} In such cases, which range from
the behavior of superfluid helium under high pressure to the phase
diagram of quantum random optimization problems, the interplay between
quantum mechanics and the otherwise classical fluctuations that lead
to vitrification can be expected to produce qualitatively novel
physical behavior.\cite{Markland11}

The theoretical investigation of quantum glasses has increased in
recent years.  Studies ranging from the investigation of quantum
effects in so-called stripe glasses,\cite{Wolynes03} quantum
spin-glasses~\cite{Cugliandolo98,Cugliandolo99,Cugliandolo01,Biroli01,Cugliandolo02,Cugliandolo04,Cugliandolo06}
and lattice models that mimic the properties of superfluid and
supersolid helium~\cite{Zamponi11} have been presented.  In this work
we instead focus on "realistic" off-lattice quantum fluids.  The
microscopic detail of our study necessitates the use of
approximations, such as mode-coupling theory (MCT)~\cite{Gotze09} and
ring-polymer molecular dynamics (RPMD),\cite{Manolopoulos04} that are
less well-justified then the methods employed in the studies of the
model systems mentioned above.  On the other hand, the approaches used
here have lead to a host of non-trivial predictions both for classical
glass-forming liquids~\cite{Gotze09} as well as a variety of quantum
liquid-state phenomena.\cite{Rabanireview05} We thus expect that the
predictions made in this work to be at least of qualitative accuracy.

The work presented here builds on our earlier report of several novel
effects that arise when glassy dynamics occurs in the quantum
regime.\cite{Markland11} In particular, both RPMD and the quantum
version of mode-coupling theory (QMCT) indicate that the dynamical
phase diagram of glassy quantum fluids is reentrant.  As a
consequence, hard-sphere quantum liquids may be forced deeper into the
glass "phase" at fixed volume fraction as quantum fluctuations
increase. This counterintuitive finding has implications not only for
liquid-state systems such as superfluid helium under pressure, but for
a broad class of quantum optimization problems as well.

In comparison to our earlier paper,\cite{Markland11} the work
presented here provides complete details for both the QMCT and the
quantum integral equations needed for generating the required
structural input.  In addition, we give a far more extensive
interpretation of the results, largely afforded by our RPMD
simulations.  Lastly, we discuss in greater detail the connection of
our results to related theoretical work.

The paper is organized as follows: In Sec.~\ref{sec:qmct} we provide
the details of the QMCT, including a description of the equations for
the density correlator and the mode coupling approximations. In
addition, we discuss the high and low temperature limit of the QMCT
and derive equations for the nonergodic parameter used to determine
the liquid-glass line. In Sec.~\ref{sec:qiet} we describe the quantum
integral equation theory used to obtain the the static input required
by QMCT. Sec.~\ref{sec:rpmd} is devoted to the RPMD method.  Results
and discussions are presented in Sec.~\ref{sec:res}. Finally, in
Sec.~\ref{sec:con} we conclude.

\section{A self-consistent quantum mode-coupling theory}
\label{sec:qmct}
The general quantity of interest is the Kubo transform~\cite{Kubo95}
of the time correlation of the collective density operator,
$\hat{\rho}_{\bf q}=\sum_{\alpha=1}^{N} \mbox{e}^{i {\bf q} \cdot {\bf
    \hat{r}}_{\alpha}}$, given by
\begin{eqnarray}
\phi_{q}(t) & = & \frac{1}{N\hbar\beta}\int_{0}^{\hbar\beta}
d\lambda\langle \hat{\rho}_{q}^{\dagger}(t+i\lambda)
\hat{\rho}_{q}(0)\rangle\nonumber \\ & \equiv & \frac{1}{N}
\left(\hat{\rho}_{q}(t)| \hat{\rho}_{q}(0)\right),
\label{eq:phiqt-def}
\end{eqnarray}
with a time evolution described by the {\em exact} quantum generalized
Langevin equation (QGLE)~\cite{Rabanireview05}
\begin{equation}
\ddot{\phi}_{q}(t) + \Omega_{q}^{2}\phi_{q}(t) + \int_{0}^{t}d\tau
M_{q}(\tau)\dot{\phi}_{q}(t-\tau) = 0,
\label{eq:qgle}
\end{equation}
In the above, we have used the notion that ${\bf \hat{r}}_{\alpha}$
stands for the position vector operator of particle $\alpha$ with a
conjugate momentum ${\bf \hat{p}}_{\alpha}$ and mass $m$, $N$ is the
total number of particles, $\beta=\frac{1}{k_{\mbox{\tiny{B}}}T}$ is
the inverse temperature and $\langle \cdots \rangle$ in
Eq.~(\ref{eq:phiqt-def}) denotes a quantum mechanical ensemble
average.  The frequency and memory terms are given by:
\begin{equation}
\Omega_{q}^{2}=\frac{q^{2}}{m \beta\phi_{q}(0)}
\label{eq:omega}
\end{equation}
and 
\begin{equation}
M_{q}(t) = \frac{\left(Q_1{\mathcal{L}^{2}}\hat{\rho}_{q} |
\mbox{e}^{i\bar{\mathcal{L}}t}|Q_1{\mathcal{L}}^{2}
\hat{\rho}_{q}\right)} {\Omega_{q}^{2}\phi_{q}(0)},
\label{eq:memory}
\end{equation}
respectively, with ${\mathcal L}=\frac{1}{\hbar}[{\hat H},\cdots]$
being the Liouvillian and $\bar{\mathcal L}=Q_{2}Q_1{\mathcal{L}}Q_1
Q_{2}$.  To derive the above equations we have defined two projection
operators (first and second order,
respectively)\cite{Gotze76a,Pathak86}
\begin{equation}
P_1=\left.|\hat{\rho}_{q}\right) \phi_{q}^{-1}(0) \left(\hat{\rho}_{q}|
\right.
\label{eq:projection}
\end{equation}
and
\begin{equation}
P_{2}= \left.|Q_1{\mathcal L} \hat{\rho}_{q} \right) \left(Q_1{\mathcal L}
\hat{\rho}_{q} |Q_1{\mathcal L}\hat{\rho}_{q} \right)^{-1}
\left(Q_1{\mathcal L}\hat{\rho}_{q}|\right.
\label{eq:projection2}
\end{equation}
with $Q_1=1-P_1$ and $Q_2=1-P_2$. $\phi_{q}(0)$ is the zero time value
of $\phi_{q}(t)$ and can be approximated by~\cite{Gotze76a}
$\frac{2S_{q}} {\beta \hbar \Delta n(\Omega_{q}) \Omega_{q}}$ where
$S_{q}$ is the static structure factor, $\Delta n(\omega) =
n(\omega)-n(-\omega)$ and $n(\omega)=\frac{1}{e^{\beta\hbar\omega}-1}$
is the Bose distribution function at temperature $T$.

\subsection{Quantum Mode-Coupling Approach}
We employ a quantum mode-coupling approach recently described by us
for quantum liquids~\cite{Kletenik11} to obtain the memory kernel
described by Eq.~(\ref{eq:memory}).  This approach is based on our
early work to describe density fluctuations and transport in quantum
liquids such as liquid {\em para}-hydrogen, {\em ortho}-deuterium, and
normal liquid
helium.\cite{Reichman01a,Rabani02a,Reichman02a,Rabani02b,Rabani02c,Rabani02d,Rabani04,Rabani05,Rabani05a}
The basic idea behind this approach is that the random force projected
correlation function, which determines the memory kernel for the
intermediate scattering function, decays at intermediate and long
times predominantly into modes which are associated with
quasi-conserved dynamical variables.  It is reasonable to assume that
the decay of the memory kernel at long times will be governed by those
modes that have the longest relaxation time.  Thus, the first
approximation made by the QMCT is to
replace the projected time evolution operator, $\mbox{e}^{i
  {\bar{\mathcal L}} t}$, by its projection onto the subspace spanned
by these slow modes.\cite{Rabanireview05} The second approximation
involves the factorization of four-point density correlations into a
product of two-point density correlation.\cite{Rabanireview05}

Following the derivation outlined by G{\"{o}}tze and L\"{u}cke (GL)
for zero temperature,\cite{Gotze76a,Gotze76b} the memory kernel at
finite temperature (in frequency space),$\tilde{M}_{q}(\omega) =
\int_{-\infty}^{\infty} dt \mbox{e}^{-i \omega t} M(q,t)$), can be
approximated by
\begin{eqnarray}
\tilde{M}_{q}(\omega) &\approx& \frac{\hbar m \beta^{2}}{4\pi\omega
  q^{2} n} \int \frac{d^{3}k}{(2\pi)^3} V_{q,k,q-k}^{2}
\int_{-\infty}^{\infty}d\omega' \omega' \\ \nonumber & & \times
(\omega-\omega') T(\omega',\omega-\omega') \tilde{\phi}_{q-k}(\omega')
\tilde{\phi}_{k} (\omega-\omega'),
\label{eq:Mqmct}
\end{eqnarray}
where $n$ is the number density,
$\tilde{\phi}_{q}(\omega)=\int_{-\infty}^{\infty} dt \mbox{e}^{i
  \omega t} \phi_{q}(t)$ is the Fourier transform of the Kubo
transform of the intermediate scattering function and
\begin{eqnarray}
T(\omega_{1},\omega_{2})
= n(-\omega_{1}) n(-\omega_{2}) - n(\omega_{1}) n(\omega_{2}).
\label{eq:T}
\end{eqnarray}
The vertex, $V_{q,k,q-k}$, is formally given by
\begin{eqnarray}
N_{q-k,k} V_{q,k,q-k} &=& \left(Q{\mathcal{L}}^{2}\hat{\rho}_{q}
|\hat{\rho}_{k}\hat{\rho}_{q-k}\right) \\ \nonumber &=&
\left(\mathcal{L}^{2}\hat{\rho}_{q}|\hat{\rho}_{k}\hat{\rho}_{q-k}\right)
- \Omega_{q}^{2} \left(\hat{\rho}_{q}| \hat{\rho}_{k}
\hat{\rho}_{q-k}\right),
\label{eq:vertex}
\end{eqnarray}
with the normalization approximated by 
\begin{eqnarray}
N_{q-k,k} &=& \left( \hat{\rho}_{q} \hat{\rho}_{q-k}| \hat{\rho}_{q-k}
\hat{\rho}_{k} \right)(0) \\ \nonumber &\approx& \hbar \beta
\int_{-\infty}^{\infty} \frac{d\omega} {\pi} \int_{-\infty}^{\infty}
\frac{d\omega'} {\pi} \frac{1} {4\omega}T(\omega',\omega-\omega')
\\ \nonumber & & \times \omega'(\omega-\omega')
\tilde{\phi}_{q-k}(\omega') \tilde{\phi}_{k}(\omega-\omega'),
\label{eq:normalization}
\end{eqnarray}
consistent with the spirit of QMCT where four-point density
correlations are factorized into a product of two-point density
correlations.\cite{Rabanireview05}

\subsection{The vertex}
The vertex in Eq.~(\ref{eq:vertex}) is difficult to compute since it
involves three-point Kubo density correlations.  A common approach
taken by classical mode-coupling theory (CMCT) is based on a
convolution approximation.\cite{Jackson62} For the Kubo transform
quantum case, a convolution-like approach is not unique. The approach
we adopt here is based on an extension of the work of GL to finite
temperatures.\cite{Gotze76a,Gotze76b} In this work, a dynamical
approximation is made to remove the dependence on Kubo transformed
structure factor in the vertex.  The assumption behind this
approximation is that the major contribution to the vertex and its
normalization comes from a characteristic frequency of the
system. Thus, we approximate $\tilde{\phi}_{q}(\omega)$ within the
vertex by
\begin{equation}
\tilde{\phi}_{q}(\omega)=\frac{2\pi S_{q}}{\beta \hbar \Delta
n(\Omega_{q}) \omega}
(\delta(\omega-\Omega_{q})-\delta(\omega+\Omega_{q})).
\label{eq:phiqw-approx}
\end{equation}
which satisfies the known sum rule $\int_{-\infty}^{\infty} d\omega
\tilde{\phi}_{q}(\omega) = \phi_{q}(0)$. Inserting this approximation
for $\tilde{\phi}_{q}(\omega)$ into the expression for $N_{q-k,k}$
given by Eq.~(\ref{eq:normalization}) yields:
\begin{equation}
N_{q-k,k} \approx \frac{2S_{q-k}S_{k}} {\hbar\beta\Delta
  n(\Omega_{q-k})\Delta n(\Omega_{k})} K(\Omega_{q-k},\Omega_{k}),
\label{eq:N-approx}
\end{equation}
where 
\begin{equation}
K(\Omega_{q-k},\Omega_{k}) = \frac{T(\Omega_{q-k},\Omega_{k})}
{\Omega_{q-k}+\Omega_{k}} + \frac{T(-\Omega_{q-k},\Omega_{k})}
{\Omega_{q-k}-\Omega_{k}}.
\label{eq:K}
\end{equation}
For $V_{q,k,q-k}$ we use the exact relations~\cite{Kletenik11}
\begin{equation}
\left(L^{2}\hat{\rho}_{q} |\hat{\rho}_{k} \hat{\rho}_{q-k}\right) =
\frac{1}{m\beta}\left(q\cdot k S_{q-k}+q\cdot(q-k)S_{k}\right).
\label{eq:gl1}
\end{equation}
and the convolution approximation
\begin{equation}
\left<\hat{\rho}_{q}^{\dagger}(t),\hat{\rho}_{k}\hat{\rho}_{q-k}\right>\approx
S_{q}S_{k}S_{q-k}
\label{eq:convGL}
\end{equation}
to obtain the approximation to the vertex:
\begin{widetext}
\begin{eqnarray}
V_{q,k,q-k} &=& \frac{\Delta n(\Omega_{q-k})\Delta n(\Omega_{k})
  C_{q,k,q-k}}{S_{q-k}S_{k}K(\Omega_{q-k},\Omega_{k})}
\left[\frac{(\Omega_{k} +\Omega_{q-k})^{2}-\Omega_{q}^{2}}
  {(\Omega_{k}+\Omega_{q-k})}\right]
\label{eq:vertexGL}
\end{eqnarray}
where
\begin{eqnarray}
C_{q,k,q-k}&=&\frac{\Omega_{q}S_{q}S_{k}S_{q-k}-\frac{\hbar \Delta
n(\Omega_{q})}{2m}\left[q\cdot kS_{q-k}+q\cdot(q-k)S_{k}\right]}
{\Omega_{q}\Delta n(\Omega_{k}+\Omega_{q-k})
-(\Omega_{k}+\Omega_{q-k})\Delta n(\Omega_{q})}
\label{eq:C}
\end{eqnarray}
\end{widetext}
The above expressions close the equation of motion
(Eq.~(\ref{eq:qgle})) and require only the static structure factor to
produce a full approximation to the time dependence of the quantum
density-density time autocorrelation function.

\subsection{High and Low temperature limits}
It may be shown that the above equations reduce to the
venerable classical mode-coupling equations in the high temperature
limit and to the GL theory as $T \rightarrow 0$.  The latter theory
produces a representation of the dispersion of superfluid helium that
is at least as accurate as the Feynman-Cohen (FC)
theory~\cite{Feynman56} at low values of $q$ and exhibits
Pitaevskii-bending of the spectrum at high $q$, unlike the FC theory.
In particular at high $T$,
\begin{eqnarray}
\lim_{\beta\rightarrow0}M_{q}(t)&\approx&\frac{k_{\rm B}T
  n}{16\pi^{3}m q^2} \int d^{3}k \left({q}\cdot k c_{k}
\right.\\ \nonumber &+& \left. {q} \cdot(q-k) c_{q-k}\right)^{2}
\phi_{q-k}(t)\phi_{k}(t),
\label{eq:Mqtclassical}
\end{eqnarray}
where $c_{q} = \frac{1}{n} \left(1-\frac{1}{S_{q}} \right)$ is the
direct correlation function.  In addition, $\phi_{q}(t)$ reduces to
the classical intermediate scattering function, $F(q,t)$ as $\beta \to
0$. This is recognized as the CMCT memory function.\cite{Gotze09}

At $T \rightarrow 0$ the equation for the memory function reduces to:
\begin{eqnarray}
\lim_{T\rightarrow0}\tilde{M}_{q}(\omega)&\approx&\frac{\hbar
  m\beta^{2}}{2 n \omega q^{2}} \int \frac{d^{3}k}{(2\pi)^{3}}
  V_{q,k,q-k}^{2} \\ \nonumber & & \times \int_{0}^{\omega}
  \frac{d\omega'}{\pi}\omega'(\omega-\omega')
  \tilde{\phi}_{q-k}(\omega')\tilde{\phi}_{k}(\omega-\omega'),
\label{eq:MqtT=0}
\end{eqnarray}
with
\begin{eqnarray}
\lim_{T\rightarrow0}V_{q,k,q-k} = & &\frac{\hbar n}{2m} (\omega_{k} +
\omega_{q-k} + \omega_{q}) \\ && \nonumber \times \left(q\cdot k c_{k}
+ q\cdot(q-k)c_{q-k}\right),
\label{eq:vertexT=0}
\end{eqnarray}
which are the $T \rightarrow 0$ equations for quantum density
fluctuations in superfluid helium first derived by
GL.\cite{Gotze76a,Gotze76b} In the above, $\omega_q=\frac{\hbar
q^2}{2mS_q}$.  We note in passing that the term $\beta^2$ appearing in
Eq.~(\ref{eq:MqtT=0}) (and not in the derivation of GL) arises from
our definition of the Kubo transform (Eq.~\ref{eq:phiqt-def}), which
includes a $\frac{1}{\hbar \beta}$, while that of GL does not. Care
must be taken applying the Kubo transform as $T \rightarrow 0$.

In the $T \rightarrow 0$ case, the entire structure of the memory
function differs greatly from that of its high temperature counterpart
and the convolution structure is lost.  Eqs.(\ref{eq:MqtT=0}) and
(\ref{eq:vertexT=0}) do not imply a memory function that is a product
of correlators at identical times.  This is a consequence of the
quantum fluctuation-dissipation theorem (QFDT) that must be
satisfied. At $T \rightarrow 0$ the function $T(\omega_q,\omega_k)$
becomes proportional to the difference of a product of step-functions
in frequency, dramatically altering the structure of the theory.  This
distinction between the low and high temperature limits has important
consequences, as discussed below.

\subsection{Nonergodic parameter}
The nonergodic parameter,
\begin{equation}
f_{q}=\frac{\phi_{q}(t\rightarrow\infty)}{\phi_{q}(0)}=\frac{\hbar
  \Delta
  n(\Omega_{q})\Omega_{q}}{2k_{B}TS_{q}}\phi_{q}(t\rightarrow\infty),
\label{eq:f}
\end{equation}
is often used to describe the ergodic to nonergodic transition as the
liquid is cooled down to the mode-coupling critical temperature
$T_c$. Above $T_c$ one finds a single solution where $f_{q}=0$ for all
values of $q$, while at $T_c$ the nonergodic parameter acquires a
finite value $f_{q} > 0$.\cite{Gotze92} It is simple to show that
$f_{q}$ must satisfy the equation:\cite{BalucaniZoppi}
\begin{equation}
\frac{f_{q}}{1-f_{q}}=\frac{1}{\Omega_{q}^{2}} M_{q}(t \rightarrow
\infty)
\label{eq:nep}
\end{equation}
The above equation for the nonergodic parameter reflects the structure
of the QGLE (Eq.~(\ref{eq:qgle})), and thus, is valid both in the
classical and quantal limits. In the former, the long time limit of
the memory kernel is given by $M_{q}(t\rightarrow\infty) \approx
\frac{k_{\rm B}T n}{16\pi^{3}m q^2} \int d^{3}k {\bar V}_{q,k,q-k}^{2}
f_{q-k} f_{k}$ with ${\bar V}_{q,k,q-k}^{2} = S_{q-k}S_{k}
\left({q}\cdot k c_{k} + {q} \cdot(q-k) c_{q-k}\right)^{2}$. The
quantum case is a bit more complicated since the structure of the
memory kernel is quite different and involves a convolution of
products of $\tilde{\phi}_{q}(\omega)$. The derivation for
$M_{q}(t\rightarrow\infty)$ is thus, based on the following expansion:
\begin{widetext}
\begin{equation}
\begin{split}
\frac{1}{\omega}T(\omega',\omega-\omega') & \omega'(\omega-\omega') =
\frac{1}{\beta \hbar} + \frac{\beta \hbar}{12}\omega'(\omega-\omega') -
\frac{(\beta \hbar)^{3}}{720}\left(\omega'^{2}-\omega'(\omega-\omega') +
(\omega-\omega')^{2}\right)+\\ & \frac{(\beta \hbar)^{5}}{30240}
\left((\omega-\omega')^{4} - \omega'(\omega-\omega')^{3} +
\omega'^{2}(\omega-\omega')^{2} -
\omega'^{3}(\omega-\omega')+\omega'^{4}\right)+O(\beta^{7}).
\end{split}
\label{eq:expand}
\end{equation}
\end{widetext}
Inserting this into the memory kernel (Eq.~(\ref{eq:Mqmct})) and
keeping the first two terms only, we obtain:
\begin{eqnarray}
\tilde{M}_{q}(\omega) &\approx& \frac{\hbar m \beta^{2}}{4\pi q^{2} n}
\int \frac{d^{3}k}{(2\pi)^3} V_{q,k,q-k}^{2}
\int_{-\infty}^{\infty}d\omega' \left(\frac{1}{\beta \hbar} + \right. \\ & &
\nonumber \left. \frac{\beta \hbar}{12}\omega' (\omega-\omega') +
\cdots\right) \tilde{\phi}_{q-k}(\omega') \tilde{\phi}_{k}
(\omega-\omega').
\label{eq:Mqmct1}
\end{eqnarray}
In the time domain, this translates to:
\begin{eqnarray}
  M_{q}(t) &\approx& \frac{\hbar m \beta^{2}}{2 q^{2} n} \int
  \frac{d^{3}k}{(2\pi)^3} V_{q,k,q-k}^{2} \\ &&
  \nonumber \times \left(\frac{1}{\beta \hbar} \phi_{q-k}(t) \phi_{k}(t) +
  \frac{\beta \hbar}{12}\dot{\phi}_{q-k}(t)\dot{\phi}_{k}(t) + \cdots\right),
\label{eq:Mqmct2}  
\end{eqnarray}
where the dot denotes a time derivative, i.e., $\dot{\phi}_{k}(t) =
\frac{d\phi_{k}(t)}{dt}$.  The other terms in the expansion of
Eq.~(\ref{eq:expand}) that have been omitted give rise to terms of the
form
\begin{equation}
  \sum_{j}a_{j}(\beta)\phi_{q-k}^{(j)}(t)\phi_{k}^{(n-j)}(t),
\end{equation}
where $a_{j}(\beta)$ are related to the expansion coefficients of
$\frac{1}{\omega}T(\omega',\omega-\omega')\omega'(\omega-\omega')$
and $\phi_{k}^{(j)}(t)=\frac{d^{j}\phi_{k}(t)}{dt^{j}}$ is the $j$'s
time derivative of $\phi_{k}(t)$.

The long time limit of the Eq.~(\ref{eq:Mqmct2}) is now given by:
\begin{eqnarray}
  M_{q}(t \rightarrow \infty) &\approx& \frac{m \beta}{2 q^{2}
    n} \int \frac{d^{3}k}{(2\pi)^3} V_{q,k,q-k}^{2} \\ & & \nonumber
  \times \phi_{q-k}(t \rightarrow \infty) \phi_{k}(t \rightarrow
  \infty),
\label{eq:Mqmct3}  
\end{eqnarray}
where all the time derivatives vanish as $t \rightarrow \infty$ even
when $\phi_{k}(t \rightarrow \infty)$ decays to a constant. Finally,
we can rewrite the above in terms of the nonergodic parameter:
\begin{eqnarray}
  M_{q}(t \rightarrow \infty) &\approx& \frac{m \beta}{2 q^{2}
    n} \int \frac{d^{3}k}{(2\pi)^3} \\ & & \nonumber \times
  V_{q,k,q-k}^{2} \phi_{q-k}(0) \phi_{k}(0) f_{q-k} f_{k}.
\label{eq:Mqmct4}
\end{eqnarray}
The above expression is strictly valid at $T \rightarrow 0$ but not at
$T=0$, since the expansion given by Eq.~(\ref{eq:expand}) is not valid
at $T=0$. The final result is similar to the classical equation,
however the vertex is given by the full quantum mechanical expression
of Eq.~(\ref{eq:vertexGL}).

\section{Quantum integral equation theory}
\label{sec:qiet}
The QMCT requires as input the static structure factor, $S_q$ and its
Kubo transform $\phi_q(0)$. Here, instead of using PIMC to generate
this input,\cite{Rabani01a} we refer to a quantum integral equation
approach, that is based on the early work of Chandler and
Richardson.\cite{Chandler84a,Chandler84b} We begin with the
Ornstein-Zernike relation applicable to quantum liquids. The quantum
system composed of $N$ particles can be mapped on a classical system
consisting of $N$ ring polymers, each polymer being composed of $P$
beads. Then, we can write the matrix RISM (reference interaction site
model~\cite{Chandler84a,Chandler84b}) equation for the classical
isomorphic system by:
\begin{equation}
h(|{\bf r}-{\bf r'}|) = w * c * w(|{\bf r}-{\bf r'}|) + n
w * c * h(|{\bf r}-{\bf r'}|),
\label{eq:QOZ}
\end{equation}
where $*$ denotes a convolution integral and as before, $n$ is the
number density.  In the above equation, $h(r)$, $w(r)$, and
$c(r)$ are the total correlation function, the self correlation
function, and direct correlation function, respectively, defined by:
\begin{equation}
\begin{split} 
h(r) = \frac{1}{\hbar\beta}\int_0^{\hbar\beta} d\lambda h(r,\lambda) \\
w(r) = \frac{1}{\hbar\beta}\int_0^{\hbar\beta} d\lambda w(r,\lambda) \\
c(r) = \frac{1}{\hbar\beta}\int_0^{\hbar\beta} d\lambda c(r,\lambda), \\
\end{split} 
\label{eq:hwc}
\end{equation}
and $h(r,\lambda)$, $w(r,\lambda)$, and $c(r,\lambda)$ are the
imaginary time total, self, and direct correlation functions,
respectively. In the classical limit Eq.~(\ref{eq:QOZ}) reduces to the
classical Ornstein-Zernike equation with $w(r)=1$. In what
follows, we will use the notation $\tilde{w}_q(\lambda)$ for the
Fourier transform of $w(r,\lambda)$, and similarly for
$\tilde{c}_q(\lambda)$ and $\tilde{h}_q(\lambda)$:
\begin{equation}
\begin{split} 
\tilde{h_q} = \frac{1}{\hbar\beta}\int_0^{\hbar\beta} d\lambda
\tilde{h}_q(\lambda) \\ \tilde{w}_q =
\frac{1}{\hbar\beta}\int_0^{\hbar\beta} d\lambda
\tilde{w}_q(\lambda) \\ \tilde{c}_q =
\frac{1}{\hbar\beta}\int_0^{\hbar\beta} d\lambda \tilde{c}_q(\lambda).
\\
\end{split} 
\label{eq:hwcq}
\end{equation}

To proceed, we refer to the mean-pair interaction approximations along
with the quadratic reference action~\cite{Chandler84a} and rewrite:
\begin{equation}
  \tilde{w}_q(\lambda) = \exp\{ -q^2 R^2(\lambda)\},
\label{eq:wqlambda}
\end{equation}
where
\begin{equation}
  R^2(\lambda) = \sum_j \frac{1-\cos(\Omega_j \lambda)}{\beta m
    \Omega_j^2 + \alpha_j},
\label{eq:R}
\end{equation}
$m$ is the particle mass, $\Omega_j=2\pi j/\hbar \beta$ is the
Matsubara frequency and $\alpha_j$ is given by:
\begin{equation}
\alpha_j = \frac{1}{6 \pi^2 \hbar \beta} \int_0^{\infty} dq
\int_0^{\hbar \beta} d\lambda q^4 \tilde{v_q} (1-\cos(\Omega_j
\lambda) \tilde{w}_q(\lambda).
\label{eq:alpha}
\end{equation}
In the above the solvent induced self-interaction is given by:
\begin{equation}
\tilde{v_q} = -\tilde{c}_q^2(n \tilde{w}_q + n^2 \tilde{h}_q).
\label{eq:v}
\end{equation}

In order to close the quantum Ornstein-Zernike equations, which in
$q$-space can be written as:
\begin{equation}
\tilde{h}_q = \tilde{w}_q \tilde{c}_q \tilde{w}_q + n
\tilde{w}_q \tilde{c}_q \tilde{h}_q,
\label{eq:QOZq}
\end{equation}
we use the Percus-Yevick (PY) closure of the form (in $r$-space):
\begin{equation}
c(r)=(h(r)+c(r)+1) (\exp(-\beta v(r)) - 1),
\label{eq:PY}
\end{equation}
where $v(r)$ is the pair interaction between two particles. The static
structure factor and its Kubo transform are then given by:
\begin{eqnarray}
  S_q = 1 + n \tilde{h}_q\\ \nonumber
  \phi_q(0) = \tilde{w}_q + n \tilde{h}_q.
\label{eq:sqfromhq}
\end{eqnarray}
In all the applications reported below we have used the approximate
relation for $\phi_{q}(0) \approx \frac{2S_{q}} {\beta \hbar \Delta
  n(\Omega_{q}) \Omega_{q}}$.

\section{Ring polymer molecular dynamics}
\label{sec:rpmd}
The RPMD approach to quantum dynamics provides an approximation to
quantum mechanical Kubo transformed correlation functions by using a
classical evolution of the imaginary time
paths~\cite{Manolopoulos04}. Consider a multidimensional system of $N$
distinguishable particles with a Hamiltonian of the form,
\begin{equation}
H = \sum_{\alpha=1}^{N} \frac{{\bf p}_{\alpha}^{2}}{2m_\alpha}+V({\bf
  r}_1,\ldots,{\bf r}_N),
\label{eq:rpmd1}
\end{equation}
where, ${\bf r}_{\alpha}$ and ${\bf p}_{\alpha}$ are the positions and
momenta of the particles and $V({\bf r}_1,\ldots,{\bf r}_N)$ is the
potential energy of the system. The RPMD approximation to the
canonical correlation function, $\tilde{c}_{AB}(t)$, for position
dependent operators $A({\bf r})$ and $B({\bf r})$ is,
\begin{eqnarray}
\tilde{c}_{AB}(t) & \simeq & \frac{1}{(2\pi\hbar)^{3NP}Z_P}\int
d^{3NP}{\bf p}\int d^{3NP}{\bf r} \nonumber \\ & & \,e^{- \beta_{P}
  H_P({\bf p},{\bf r})}A_P({\bf r})B_P({\bf r}_t),
\label{eq:cab}
\end{eqnarray}
where
\begin{equation}
Z_{P} = \frac{1}{(2\pi\hbar)^{3NP}}\int d^{3NP}{\bf p}\int d^{3NP}{\bf
  r}~ \,e^{-\beta_{P} H_{P}({\bf p},{\bf r})},
\end{equation}
and $\beta_{P} =\beta/P$. $H_P({\bf p},{\bf r})$ is the classical
Hamiltonian of the $N$ particle $P$ bead ring polymers with the
external potential of $V({\bf r}_1,\ldots, {\bf r}_N)$ acting on each
bead,
\begin{eqnarray}
H_P({\bf p},{\bf r}) & = & \sum_{\alpha=1}^{N} \sum_{k=1}^{P}
\left(\frac{({\bf
    p}_{\alpha}^{(k)})^2}{2m_i}+\frac{1}{2}m_{\alpha}\omega_P^2 ({\bf
  r}_{\alpha}^{(k)}-{\bf r}_{\alpha}^{(k+1)})^2\right) \nonumber \\ &&
+\sum_{k=1}^{P} V({\bf r}_1^{(k)},\ldots,{\bf r}_N^{(k)}),
\label{eq:hp}
\end{eqnarray}
where $\omega_{P}=1/\beta \hbar$ and the cyclic boundary condition
${\bf r}_{\alpha}^{(P+1)} \equiv {\bf r}_{\alpha}^{(1)}$ applies. The
time-evolved coordinates ${\bf r}_t\equiv {\bf r}_t({\bf p},{\bf r})$
in Eq.~(\ref{eq:cab}) are obtained from the classical dynamics
generated from this Hamiltonian and the operators $A_P({\bf r})$ and
$B_P({\bf r}_t)$ are evaluated by averaging over the beads of the ring
polymer at times $0$ and $t$ respectively,
\begin{equation}
A_P({\bf r}) = \frac{1}{P}\sum_{k=1}^P A({\bf r}_1^{(k)},\ldots,{\bf r}_N^{(k)}),
\label{eq:rpmd11}
\end{equation}
\begin{equation}
B_P({\bf r}) = \frac{1}{P}\sum_{k=1}^P B({\bf r}_1^{(k)},\ldots,{\bf r}_N^{(k)}).
\label{eq:rpmd12}
\end{equation}
The RPMD method has previously been used to study a diverse selection
of multidimensional systems including proton transfer between organic
molecules,\cite{Collepardo-Guevara2008} diffusion in and inelastic
neutron scattering from liquid
para-hydrogen,\cite{Miller2005,Craig2006} diffusion of light atoms in
liquid water,\cite{Markland2008} and gas phase reactions such as that
between methane and hydrogen.\cite{Suleimanov2011} In all cases RPMD
has been able to capture the dominant quantum mechanical effects in
the dynamics and provide good agreement with the available
experimental or exact results. RPMD has also been applied to look at
deep tunneling of Muonium and Hydrogen atoms in
ice~\cite{Markland2008} and in this regime has been shown to be
related to semi-classical Instanton theory.\cite{Richardson09}

\section{Simulations Details}
\label{sec:sim}

\begin{figure*}[t]
\includegraphics[width=16cm]{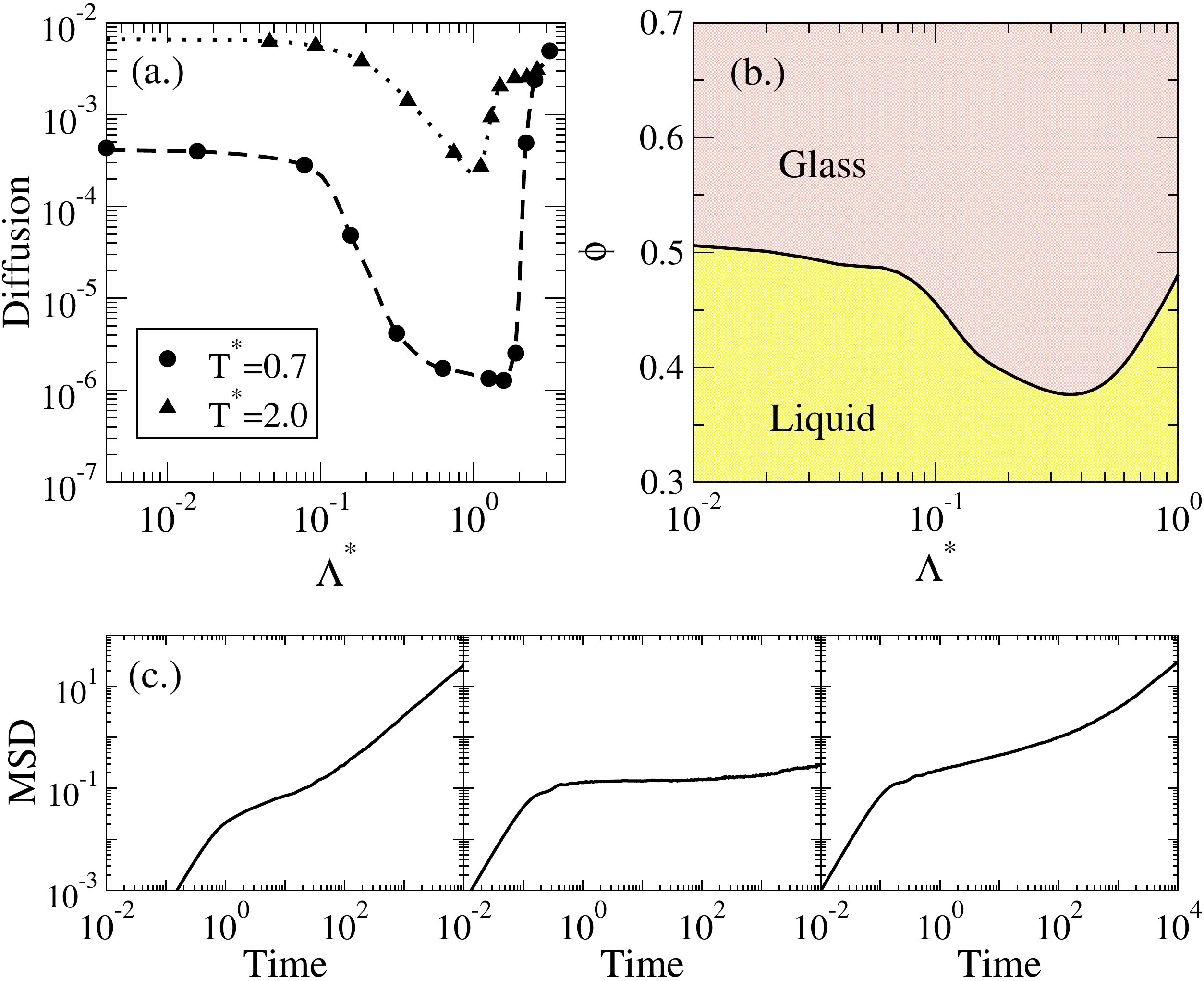}
\caption{Panel~(a.): The diffusion constant of particles of type
  A as a function of the quantumness, $\Lambda^{*}$, obtained from the
  RPMD simulations for a quantum Kob-Anderson LJ binary mixture for
  two temperatures. Panel~(b.): Dynamic phase diagram (volume
  fraction versus quantumness) calculated from the QMCT for a
  hard-sphere fluid. Panel~(c.): The mean square displacement of A
  particles as obtained from the RPMD simulations for the classical
  case (left frame, $\Lambda^{*}=0$), the trapped regime (middle
  frame, $\Lambda^{*}=1.125$), and the regime governed by strong
  quantum fluctuations (right frame, $\Lambda^{*}=1.1325$).}
\label{fig:phase}
\end{figure*}

The quantum mode coupling theory requires as input the static
structure factor and its Kubo transform. In the present study, we used
a single component hard sphere (HS) model to generate this input
within the frame work of the integral equation approach described
above. Using the PY closure, the system remains disordered even at
very high volume fraction, thus providing a simple model to explore
the quantum glass transition.  The integral equations
(\ref{eq:hwcq})-(\ref{eq:alpha}) were solved self-consistently. A
simple trapezoidal integration scheme over the imaginary time axis was
employed, with $P=400$ slices (we have checked convergence of the
static input with respect to $P$). Here, $P$ is analogous to the
number of beads in the RPMD approach.  For the HS system, it can be
shown that the quantum mode coupling equations scale with the ratio of
the de Broglie thermal wavelength to the particle size,
$\Lambda^{*}=\sqrt{\beta \hbar^2/m \sigma^2}$. Thus, to change the
quantumness, one can either change $\hbar$, or the mass, or the
temperature. For the QMCT results shown below, we have varied the
temperature to reflect a change in $\Lambda^{*}$. We note that the
temperature has no effect on the static structure factor in the
classical case.

\begin{table}
\begin{tabular*}{0.5\textwidth}{@{\extracolsep{\fill}}ccc} 
\hline\hline 
\multicolumn{1}{c}{Parameter} & \multicolumn{1}{c}{LJ units} & \multicolumn{1}{c}{Atomic Units} \\ 
\hline
$\epsilon_{AA}$ & 1  &  3.8x10$^{-4}$\\
$\epsilon_{BB}$ & 0.5  &  1.9x10$^{-4}$\\
$\epsilon_{AB}$ & 1.5  &  5.7x10$^{-4}$\\
$\sigma_{AA}$   & 1  & 6.43 \\
$\sigma_{BB}$   & 0.88  & 5.65 \\
$\sigma_{AB}$   & 0.8  & 5.14 \\
Mass$_{A}$      & 1  & 3646 \\ 
Mass$_{B}$      & 1  & 3646 \\\hline \hline
\end{tabular*}
\caption[]{Parameters used in our RPMD simulations on the Andersen-Kob
  Lennard-Jones glass forming system.}
\label{ta:param}
\end{table}

We performed RPMD simulations on the Kob-Andersen glass forming
system,~\cite{Kob95a,Kob95b} a binary LJ fluid, because the HS system
investigated above by means of QMCT crystallizes on the timescale of
the RPMD simulations.  Each simulation consisted of $1000$ particles,
$800$ of type A and $200$ of type B in a cubic box of length
$9.4\sigma_{AA}$. The LJ parameters are given in
Table~\ref{ta:param}.The equations of motion were integrated using a
time step of 0.005 in Lennard-Jones (LJ) units using the normal mode
integration scheme of Ref.~\onlinecite{Ceriotti2010}. The number of
beads, $P$, used was given by the formula,
\begin{equation}
P = \frac{11.2 \hbar}{T^*}.
\end{equation}
This choice gives good convergence for all the regimes studied.
Initial configurations were generated by annealing from a temperature
$T^{*}=5.0$ to the target temperature over a period of $2\times 10^6$
time-steps. From these initial configurations we ran a further
$2\times 10^5$ steps of equilibration using a targeted Langevin
equation normal mode thermostatting scheme.\cite{Ceriotti2010} This
was followed by microcanonical dynamics for $2\times 10^6$ steps
during which the results were collected. The quantum effect,
$\Lambda^{*}$, was varied by changing the parameter $\hbar$. Five
simulations were run for each temperature and value of $\hbar$ and the
results averaged.

\section{Results}
\label{sec:res}
Fig.~\ref{fig:phase} shows the results obtained from our QMCT
treatment of hard spheres and RPMD simulations of the KA binary LJ
fluid as the size of quantum fluctuations in the system are
varied.\cite{Markland11} Both of these systems have previously been
shown to exhibit all of the features of glassy behavior present in
more complex fluids. In panel~(b.) we show the liquid-glass dynamic
phase diagram that is obtained from the QMCT calculation. The phase
boundary is defined as the point where the solution of equations
Eqs.~(\ref{eq:f}), (\ref{eq:nep}) and (\ref{eq:Mqmct4}) leads to a
finite value for the nonergodic parameters, $f_q$. At this point QMCT
predicts that the system will never fully relax on any time-scale at
the given packing fraction. For the RPMD calculations, which are based
on the evolution of semi-classical trajectories, we instead show the
effect of quantum fluctuations on the diffusion coefficient of the
particles at two different temperatures ($T^{*}=2.0$ and $0.7$) as the
classical glass transition temperature of the system is approached
($T^{*} \approx 0.45$) in panel~(a.) of Fig. \ref{fig:phase}. Since
the mean square displacement of the particles in the ring polymer
trajectories show a caging regime (see the panel~(c.) of
Fig.~\ref{fig:phase}), the diffusion constant was extracted from the
long time slope of the mean-square displacement where the diffusive
regime had been reached. The size of the quantum fluctuations were
controlled by varying $\Lambda^{*}$, the ratio of the de Broglie
thermal wavelength to the particle size which controls the scale of
quantum behavior.

Comparing the RPMD results in panel~(a.) and QMCT results in
panel~(b.) of Fig.~\ref{fig:phase}, a remarkably consistent picture
emerges from these two different approaches to quantum dynamics and
glass forming systems. In the classical limit ($\Lambda^{*}
\rightarrow 0$) RPMD reduces to classical mechanics and QMCT to
classical MCT. As small quantum fluctuations are initially introduced,
little difference is observed in either the RPMD diffusion coefficient
or QMCT liquid-glass line. However, as $\Lambda^{*}$ is increased
beyond 0.1, quantum effects are at first found to promote and then
inhibit glass formation. In the case of RPMD, this is characterized by
a decrease of nearly three orders of magnitude in the diffusion
coefficient, and for QMCT, a 20 \% fall in the packing fraction
required for vitrification. When the thermal wavelength is increased
further and becomes on the order of the particle size, the diffusion
coefficient in the quantum system exceeds that observed in the
classical limit.  In addition the RPMD simulations at $T^{*}=0.7$ and
$2.0$ indicate that size of the re-entrance becomes much larger as the
glass transition temperature is approached.  Moreover, there is a hint
of an interesting effect where, at high values of $\Lambda^{*}$, the
diffusion coefficient at lower temperature exceeds that at the higher
temperature. We will return to this point later.

\begin{figure}[t]
\includegraphics[width=8cm]{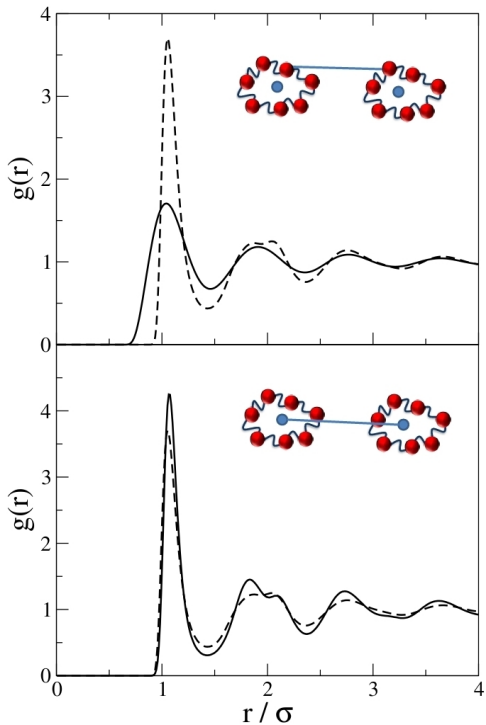}
\caption{The bead (upper panel) and centroid (lower panel) radial
  distribution functions of A particles for a classical
  ($\Lambda^{*}=0$, dashed) and trapped quantum ($\Lambda^{*}=0.75$,
  solid) regime. The bead distribution suggests less order in the
  trapped regime compared to a classical simulation while the centroid
  structure shows an increase in order.}
 \label{fig:gr}
\end{figure}

Since both MCT and our new QMCT approach use the structure factor as
input it is instructive to see if the dynamical reentrance is hinted
at in this property. Fig.~\ref{fig:gr} shows the radial distribution
function (RDF), which is the spatial Fourier transform of the
structure factor, that has been calculated from the RPMD simulations
of the KALJ fluid. For static equilibrium properties such as the RDF,
RPMD gives numerically exact results since it reduces to the path
integral molecular dynamics approach.\cite{Parrinello1984} The true
(observable) quantum RDF is determined by the ring polymer bead
correlations and is shown in the top panel for both the classical
limit ($\Lambda^{*}=0$) and for a trapped regime
($\Lambda^{*}=0.75$). As quantum effects are introduced the RDF
exhibits a broadening of the peaks due to the increasing uncertainty
in the particle positions which acts to smear out the pair
structure. Throughout the entire range of $\Lambda^{*}$ studied the
structure is observed simply to broaden systematically with
$\Lambda^{*}$ and thus, there is no indication of the observed
dynamical reentrance in the RDF.

In the bottom panel we show the centroid RDF in which the centers of
the imaginary time paths, rather than the bead positions, were used to
compute the RDF. In the classical limit all beads collapse to a single
point and hence both ways of calculating the RDF become
identical. However as quantum fluctuations are increased the beads
spread further from the center of the polymer and hence the centroid
structure offers a different view into the structure of the quantum
liquid. Upon examining the centroid RDF in Fig ~\ref{fig:gr} one sees
the opposite trend upon increasing quantum fluctuations to that
observed in the bead RDF, i.e. weak quantum fluctuations lead to a
more structured centroid RDF which one would associate with more
glassy dynamics. As quantum fluctuations further increase this trend
reverses (data not shown). Hence the centroid pair distribution
function, which is not an experimental observable, appears to grossly
mimic the dynamical correlations observed in both the QMCT and RPMD
calculations.  This is not entirely surprising, because one expects
that the centroid molecular dynamics (CMD) method,\cite{Voth96r} an
approach similar to RPMD, will also capture the reentrance.  Since CMD
is an effective classical dynamics on the many-body centroid potential
and since there are situations where the many-body centroid potential
can be approximated by a sum of pair-wise potentials given by
$-k_{\tiny B}T \log g(r)$,\cite{Voth96,Voth04} such static
correlations in the centroid RDF must be evident if CMD is to
reproduce the same phenomenology as predicted by QMCT and RPMD.  This
fact suggest that a strictly classical MCT calculation that uses a
static structure factor constructed from the centroid correlations
might be a good proxy for the full QMCT calculation.  It should be
noted that the full QMCT only uses the observable structure factor and
thus one role played by the quantum vertex function is to effectively
convert the bead correlations to centroid ones via the quantum
fluctuation-dissipation theorem. The fact that the quantum vertex
involves frequency convolutions while the classical version does not
suggests, however, that there must be some distinction between a
classical MCT calculation with centroid correlations and the full
QMCT.

\begin{figure}[t]
\includegraphics[width=8cm]{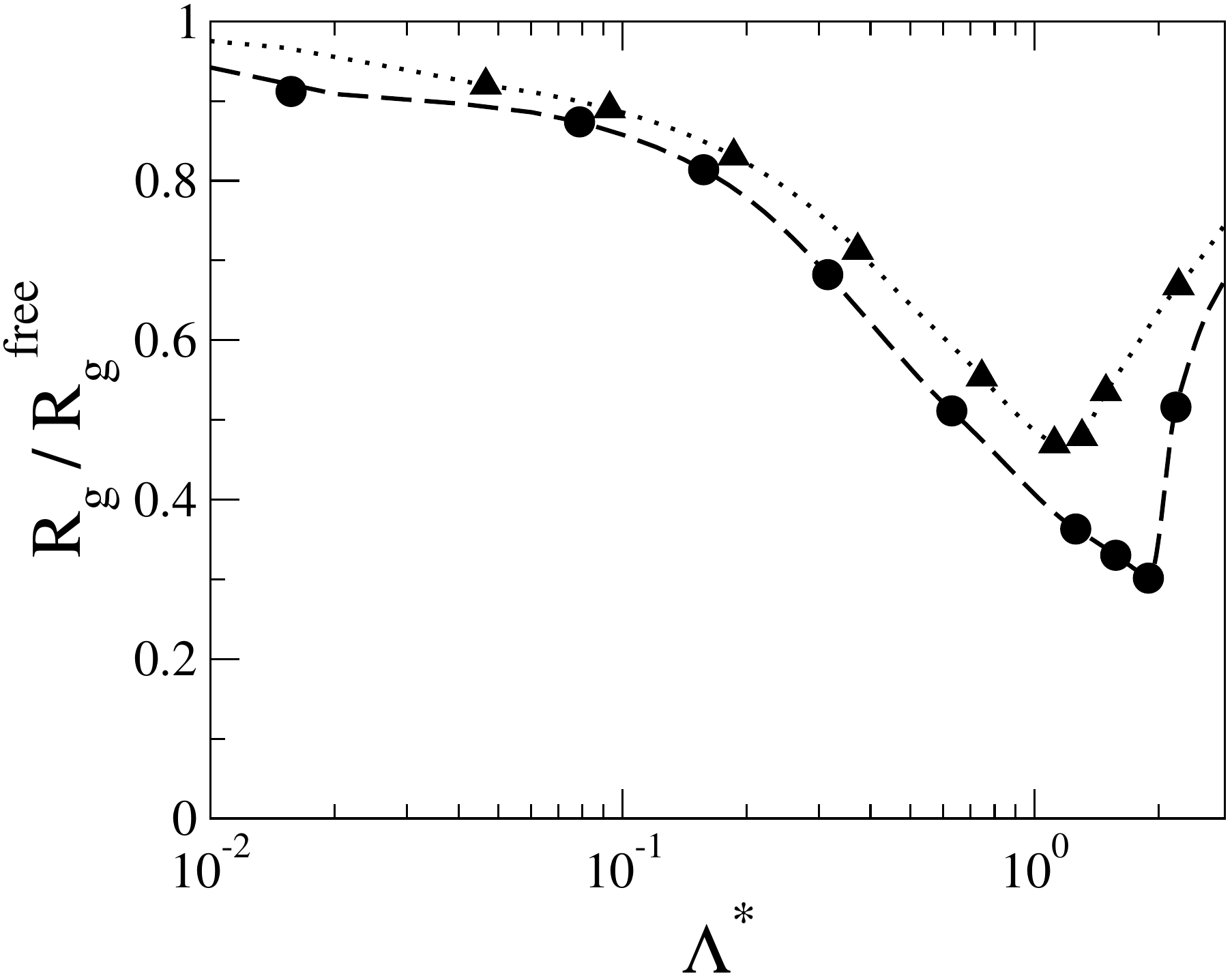}
\caption{Root-mean-square of the radius of gyration of A particles as
  a function of $\Lambda^{*}$ obtained from the RPMD simulations for a
  quantum Kob-Anderson LJ binary mixture for two temperatures. The
  radius of gyration is defined as the average distance of the
  replicas from the polymer center.  The results are plotted for
  temperatures $T^*=0.7$ (circles with dashed lines) and $T^*=2.0$
  (triangles with dotted lines).}
\label{fig:rg}
\end{figure}

So what is the origin of the reentrance? For this we turn to the RPMD
trajectories to provide a physically insightful picture. Since this
approximation maps the dynamics of a quantum mechanical particles onto
that of a system of classical ring polymers, we can initially
interpret the results in the language of the diffusion of classical
polymers. In doing so we are careful to note that each bead on a given
polymer only interacts with the bead on another ring polymer
corresponding to the same imaginary time slice, a point which we will
return to later in this section. In the non-interacting limit, the
free ring polymer radius of gyration is directly proportional to the
thermal deBroglie wavelength of the quantum particle. Hence,
increasing $\Lambda^{*}$ allows the ring polymer representing each
quantum particle to spread out. The average radius of gyration of each
quantum particle in the interacting KALJ system is a static property
which can be calculated exactly from RPMD simulations. In
Fig.~\ref{fig:rg} we plot the average radius of gyration of each ring
polymer relative to the value in the free limit.  The dependence of
this ratio on $\Lambda^*$ mimics the dependence of the diffusion
coefficient on $\Lambda^*$ shown in Fig.~\ref{fig:phase}. The decrease
in this ratio when reentrance is observed suggests a correlation
between the localization of the quantum particle and the increase in
the glassiness of the system.  As quantum fluctuations are increased
from $\Lambda^{*} < 0.1$, the effective diameter of the quantum
particles differ little from $\sigma$ so that they can still fit into
the thermally accessible space, their radius of gyration is still well
approximated by $R_g^\text{free}$, and little change in the dynamics
is observed. However, once $\Lambda^{*}$ exceeds $0.1$ there is not
enough free space for the free ring polymers to further expand and
crowding due to the surrounding solvent cage causes the radius of
gyration to decrease from its free particle value.

\begin{figure}[t]
\includegraphics[width=8cm]{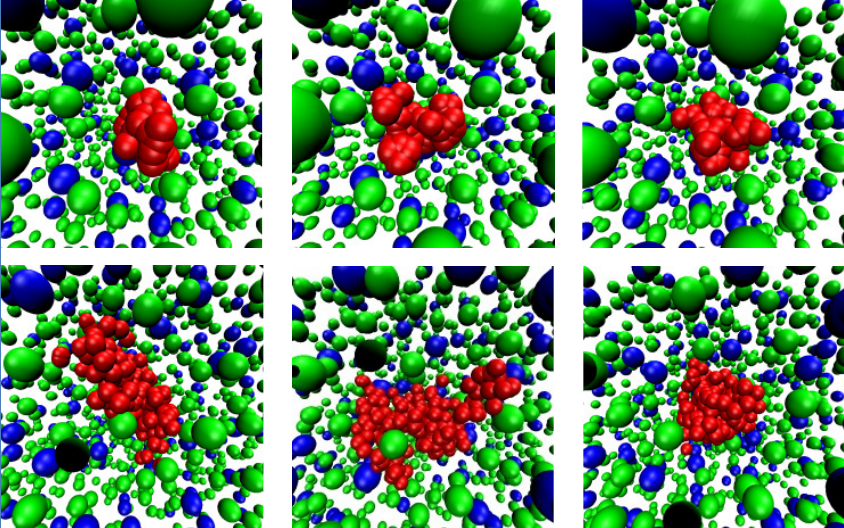}
\caption{A series of snapshots taken from simulations at
  $\Lambda^{*}=1.125$ (upper panels) and $\Lambda^{*}=1.3125$ (lower
  panels) with $T^{*}=0.7$. For clarity the full imaginary time path
  (colored red) is only shown for one particle of type A with all
  others represented by their centroids.  The centroids for the other
  particles of types A and B are colored green and blue, respectively.
  The upper panels depict configurations which reside in the trapped
  regime where the ring polymer is essentially localized in one cavity
  cage whereas in the tunneling regime (lower panels) it is frequently
  spread across two or more cavities in the liquid resulting in more
  facile motion.}
\label{fig:traj}
\end{figure}

In the upper panel of Fig.~\ref{fig:traj}, we show typical
configurations of a RPMD trajectory in the regime where the particle
is localized in a cavity. The particle is confined by its surrounding
neighbors, thus giving rise to an increase in its quantum kinetic
energy. For diffusion to occur, particles must push past each other,
causing further localization and incurring an even greater increase in
their kinetic energies. This energy penalty to motion leads to slower
dynamics. As $\Lambda^{*}$ is further increased, a tipping point is
reached when the thermal wavelength becomes comparable to the particle
size, $\Lambda^{*} \approx 1$. At this point, the cost of localization
becomes so large that the induced quantum kinetic energy enables the
crossing of barriers between cavities, leading to a rise in the radius
of gyration and facilitating diffusion. This can be seen in the
representative snapshots of a RPMD trajectory shown in the lower
panels of Fig.~\ref{fig:traj} in which the particle is delocalized
across two cavities. Accordingly, the radius of gyration recovers with
a corresponding increase in diffusion coefficient and diminishing of
the caging regime. This can be likened to a ``lakes to oceans''
percolation transition, in which the caging regime reflects
frustration of the quantum particle in the classical potential, a
frustration which is reduced when the kinetic energy of confinement
essentially floods the barriers and allows the particle to traverse
the region between adjacent potential energy minima.

Reentrant effects in quantum systems have also been observed in the
diffusion of electrons in a sea of classical random
blockers~\cite{Leung1994} as well as in model
systems.\cite{Wolynes03,Zamponi11} In the former case the problem can
be exactly mapped onto the diffusion of a classical ring
polymer. However, in our case, while the expression ``ring polymer''
is used to describe the isomorphism arising from the imaginary time
path integral representation described in
Eqs.~(\ref{eq:rpmd1})-(\ref{eq:hp}), it is not simply that of a system
of true harmonic ring polymers. This is because each bead of a polymer
only interacts with its corresponding bead at the same imaginary time
on the polymer representing another particle and not with any other
beads on that particle.  One might therefore expect that in systems
with strong interactions it might be advantageous for the polymers to
correlate their beads so as to minimize repulsion in exchange for a
loss in entropy.  To investigate this, we define vectors ${\bf
  R}^{k}_{\alpha} ={\bf r}^{(k)}_{\alpha}-{\bf r}_{\alpha}^{c}$, which
represent the position of the bead at imaginary time $k$ on ring
polymer $\alpha$ relative to the position of the centroid (${\bf
  r}_{\alpha}^c=(1/P)\sum_{k=1}^P{\bf r}_{\alpha}^{(k)})$, and we
define the angle between vectors ${\bf R}^k_{\alpha}$ and ${\bf
  R}^{(k)}_{\beta}$ as,
\begin{equation}
  \cos\theta^{(k)}_{\alpha,\beta} = \frac{{\bf R}^{(k)}_{\alpha} \cdot {\bf R}^{(k)}_{\beta}}{\mid {\bf R}^{(k)}_{\alpha} \mid \mid  {\bf R}^{(k)}_{\beta} \mid}.
\end{equation}
This function, $\cos\theta^{(k)}_{\alpha,\beta}$, will have a value
of $-1$ if the $k$-th beads on polymers  $\alpha$ and $\beta$ are aligned
perfectly away from each other and $+1$ if the beads are aligned
towards each other. Since any correlation between the beads on two
different particles is likely to be more pronounced at short distances
where interactions are stronger we plot the correlation function
$C(r)$,
\begin{equation}
  C(r) = \langle \frac{1}{N}\sum_{\alpha>\beta}\frac{1}{P}
  \sum_{k=1}^P\cos\theta^{(k)}_{\alpha,\beta} \delta(r-|{\bf
    r}_{\alpha,\beta}^c|)\rangle,
\label{eq:cr}
\end{equation}
as a function of the distance between the centroids of two ring
polymers.

\begin{figure}[t]
\includegraphics[width=8cm]{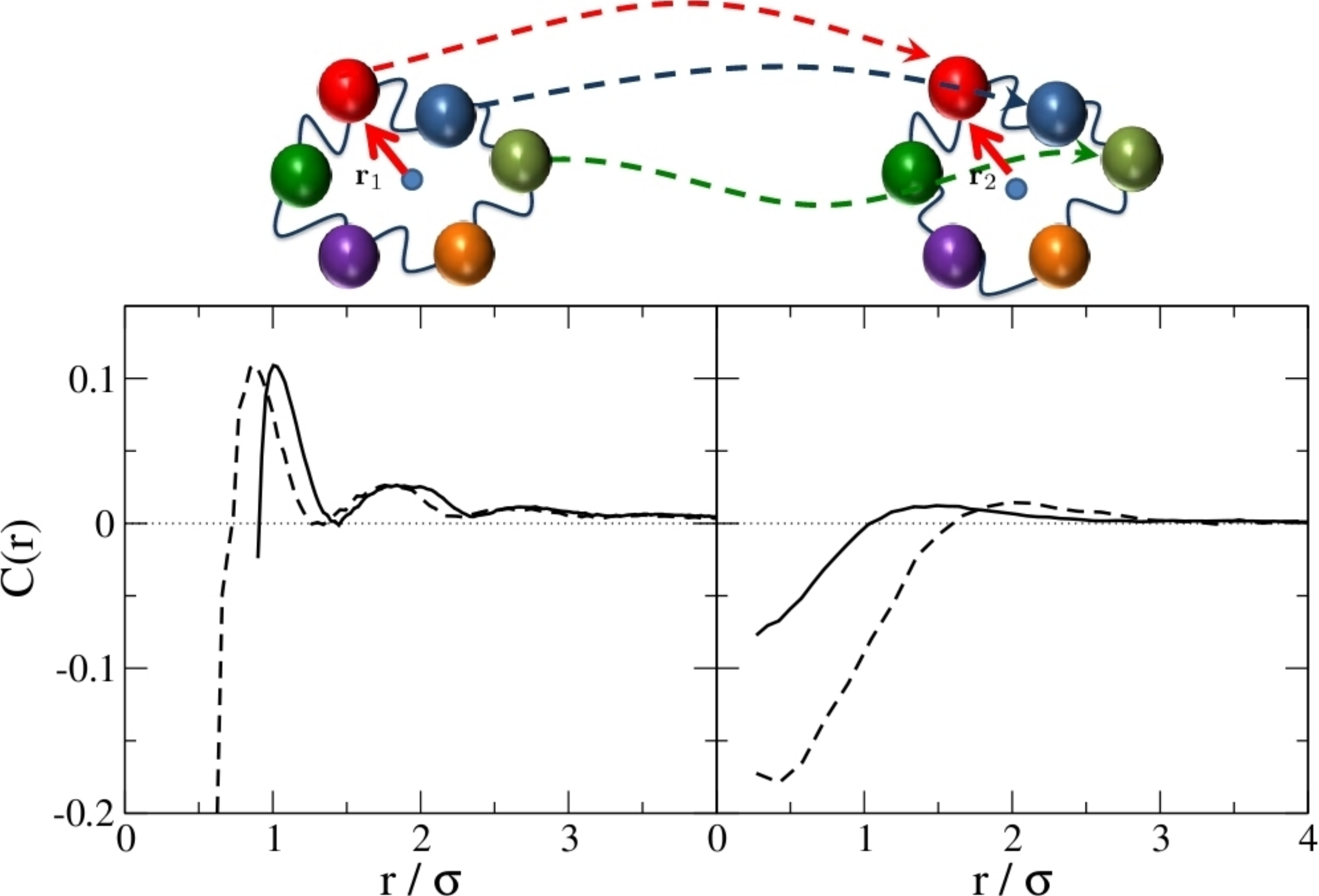}
\caption{The bead vector correlation (see Eq.~(\ref{eq:cr})) for a
  trapped regime with $\Lambda^{*}=0.75$ (left panel) and regime where
  quantum fluctuations are pronounced with $\Lambda^{*}=1.3125$ (right
  panel). The solid lines represent the bead vector correlations
  between A particles and the dashed ones those between B particles.
  In both cases $T^{*}=0.7$.  In the trapped regime the ring polymer
  beads show a large positive correlation around $r=\sigma$ which
  results in a large repulsion when the particles attempt to move past
  each other. In the other regime the beads align such that the
  correlation is largely negative which facilitates particle motion.}
\label{fig:c}
\end{figure}

The function $C(r)$ is shown in Fig.~\ref{fig:c} for $\Lambda^{*}=0.75
$, which corresponds to the trapped regime, and for $\Lambda= 1.3125$,
which corresponds to the strong quantum fluctuation regime. For $r\leq
\sigma$ $C(r)$ is negative in both cases. At these distances the
potential between particles is strongly repulsive and hence for
polymers to approach this close their beads for the same imaginary
time must avoid each other.  However for $r\approx \sigma$, $C(r)$
corresponding to the lower value of $\Lambda^{*}$, the correlation
becomes positive. At this distance the pair potential is attractive
and hence the energy of the system is lowered if the polymer arranges
its beads such that they are aligned on the same side of the
respective ring-polymers.  However at the higher value of
$\Lambda^{*}$, the entropic cost of such an ordering outweighs the
energetic benefit, and hence $C(r)$ is negative. This coincides with
the change between the dynamical regimes of quantum trapping and
strong fluctuations because, for diffusion to occur, particles must
move past each other. This regime corresponds to enhanced tunneling.
In the case of low $\Lambda^{*}$ the beads of the polymer in the first
coordination shell at $r=\sigma$ are largely aligned such that pushing
them together induces a larger repulsion than if no such correlation
existed. This increases the barrier to diffusion in this regime.

One natural question that arises from this interpretation of our
results is what occurs if the RPMD calculations are carried out at
constant pressure rather than constant volume. The analogous QMCT
calculations are constant volume calculations and indeed, as far as
the hard-sphere control variables of volume fraction and $\Lambda^{*}$
are concerned, this question is irrelevant.  The pressure varies as
the volume fraction, which can then just be rescaled to yield results
identical to those presented in the panel~(b.) of
Fig.~\ref{fig:phase}.  However from the standpoint of thermal
variation, e.g. the variation of diffusion at fixed temperature while
varying $\Lambda^{*}$ (see panel~(a.) of Fig.~\ref{fig:phase}), this
question needs to be addressed.  A natural expectation is that at
constant pressure the reentrant effect will be mitigated or destroyed
as the system can now adjust its volume as a natural response to the
buildup of local pressure created by the ``swelling'' of the ring
polymer. However, some aspects of this effect have been observed, for
example, in analogous reentrant-like effects seen in
Ref.~\onlinecite{Markland2008}, where quantization of a single species
in a constant pressure classical bath produces a reduction of the
effective diffusion constant. More generally the values of
$\Lambda^{*}$ for which the slowing of the liquid is observed are
highly realizable in room temperature systems. The thermal wavelength
of hydrogen at 300 K is 1.0 $\AA$ and hence the region of quantum
slowing corresponds to diffusion in a medium with particles of radius
2 to 5 $\AA$. Such a slow down is evident in experimental measurements
of the diffusion of hydrogen in non-glassy media such as water and
palladium.~\cite{Markland2008,Wipf1997}

We have carried out RPMD simulations of the binary glass-forming
system at constant pressure, and indeed found at least a strong
mitigation of the reentrant effect.  Currently our statistics are not
sufficient to make definitive statements about dynamical behavior in
these systems, and thus these results will be reported in a future
publication.  Regardless, it is clear that constant volume (confined)
systems will exhibit a strong enhancement of the effects reported
here. Further it should be mentioned that a similar reentrance is seen
in lattice models of quantum glasses where the concept of swelling of
imaginary time paths cannot be invoked to explain reentrant
relaxation.~\cite{Wolynes03,Zamponi11}

On a final note, a subtle feature of the RPMD results of
Fig.~\ref{fig:phase} (panel~(a.)) should be mentioned. At very large
values of $\Lambda^{*}$ the isothermal diffusion curves appear to
cross.  While the effect is quite small, this crossing would imply a
reentrance of a different sort, namely a ``melting by cooling''
mechanism.  This type of reentrance, distinct from that discussed for
the bulk of this work, is similar to that discussed in
Ref.~\onlinecite{Wolynes03}. It should be noted, however, that the
$\Lambda^{*}$ values here are large enough that particle statistics
cannot be neglected in the simulation of a realistic quantum fluid,
and the inclusion of such features may obviate this effect.

\section{Concluding remarks}
\label{sec:con}
In this work we have presented a self-contained discussion of
predictions for quantum glasses made by QMCT and RPMD. The predictions
of these two distinct, albeit highly approximate, theories appear to
be in harmony with each other.  Both predict a strong reentrance in
the relaxation of quantum supercooled liquids, namely that weak
quantum fluctuations actually serve to push the system closer towards
the glass transition.  This seemingly paradoxical effect has also been
noted in lattice models of quantum glasses and in models of quantum
optimization.  Indeed, one interesting aspect of our work is that it
suggests that typical quantum annealing protocols should generically
have regions of parameter space where they are in fact less efficient
than their classical counterparts.

Future work will be directed towards the inclusion of bosonic
statistics into the formulation of QMCT so that an investigation of
the putative superglass may be carried out in a microscopic manner.
In addition, it would be interesting to investigate more complex
liquids such as confined supercooled water to see if quantum effects
which may manifest at high temperatures lead to novel dynamical
relaxation patterns.  These topics will be reserved for the future.

\section{Acknowledgments}
The authors acknowledge Francesco Zamponi for useful discussions.  KM
acknowledges support from Kakenhi grant No. 21015001 and 2154016.  BJB
acknowledges support from NSF grant No. CHE-0910943. DRR would like to
thank the NSF through grant No. CHE-0719089 for support.  ER and DRR
thank the US-Israel Binational Science Foundation for support.

\bibliographystyle{aip}
\bibliography{qglass}

\begin{thebibliography}{10}

\bibitem{Stillinger01}
P.~G. Debenedetti and F.~H. Stillinger,
\newblock Nature {\bf 410}, 259 (2001).

\bibitem{Berthier05}
L.~Berthier et~al.,
\newblock Science {\bf 310}, 1797 (2005).

\bibitem{Biroli09}
G.~Biroli, J.~P. Bouchaud, A.~Cavagna, T.~S. Grigera, and P.~Verrocchio,
\newblock Nature Phys. {\bf 4}, 771 (2008).

\bibitem{Chandler09}
L.~O. Hedges, R.~L. Jack, J.~P. Garrahan, and D.~Chandler,
\newblock Science {\bf 323}, 1309 (2009).

\bibitem{Angell96}
M.~D. Ediger, C.~A. Angell, and S.~R. Nagel,
\newblock J. Phys. Chem. {\bf 100}, 13200 (1996).

\bibitem{Wu91}
W.~H. Wu, B.~Ellman, T.~F. Rosenbaum, G.~Aeppli, and D.~H. Reich,
\newblock Phys. Rev. Lett. {\bf 67}, 2076 (1991).

\bibitem{Imry09}
A.~Amir, Y.~Oreg, and Y.~Imry,
\newblock Phys. Rev. Lett. {\bf 103}, 126403 (2009).

\bibitem{Markland11}
T.~E. Markland et~al.,
\newblock Nature Phys. {\bf 7}, 134 (2011).

\bibitem{Wolynes03}
H.~Westfahl, J.~Schmalian, and P.~G. Wolynes,
\newblock Phys. Rev. B {\bf 68}, 134203 (2003).

\bibitem{Cugliandolo98}
L.~Cugliandolo and G.~Lozano,
\newblock Phys. Rev. Lett. {\bf 80}, 4979 (1998).

\bibitem{Cugliandolo99}
L.~Cugliandolo and G.~Lozano,
\newblock Phys. Rev. B {\bf 59}, 915 (1999).

\bibitem{Cugliandolo01}
L.~Cugliandolo, D.~Grempel, and C.~Santos,
\newblock Phys. Rev. B {\bf 64}, 014403 (2001).

\bibitem{Biroli01}
G.~Biroli and L.~Cugliandolo,
\newblock Phys. Rev. B {\bf 64}, 014206 (2001).

\bibitem{Cugliandolo02}
L.~Cugliandolo, D.~Grempel, G.~Lozano, H.~Lozza, and C.~Santos,
\newblock Phys. Rev. B {\bf 66}, 014444.

\bibitem{Cugliandolo04}
L.~Cugliandolo, D.~Grempel, G.~Lozano, and H.~Lozza,
\newblock Phys. Rev. B {\bf 70}, 024422 (2004).

\bibitem{Cugliandolo06}
L.~F. Cugliandolo,
\newblock International J. Mod. Phys. B {\bf 20}, 2795 (2006).

\bibitem{Zamponi11}
L.~Foini, G.~Semerjian, and F.~Zamponi,
\newblock Phys. Rev. B {\bf 83}, 094513 (2011).

\bibitem{Gotze09}
W.~G{\"{o}}tze,
\newblock {\em Complex Dynamics of Glass-Forming Liquids: A Mode-Coupling
  Theory},
\newblock Oxford University Press, Oxford, 2009.

\bibitem{Manolopoulos04}
I.~R. Craig and D.~E. Manolopoulos,
\newblock J. Chem. Phys. {\bf 121}, 3368 (2004).

\bibitem{Rabanireview05}
E.~Rabani and D.~R. Reichman,
\newblock Ann. Rev. Phys. Chem. {\bf 56}, 157 (2005).

\bibitem{Kubo95}
R.~Kubo, M.~Toda, and N.~Hashitsume,
\newblock {\em Statistical Physics II},
\newblock Solid State Sciences, Springer, Berlin, 2nd edition, 1995.

\bibitem{Gotze76a}
W.~G{\"{o}}tze and M.~L{\"{u}}cke,
\newblock Phys. Rev. B {\bf 13}, 3822 (1976).

\bibitem{Pathak86}
J.~S. Thakur and K.~N. Pathak,
\newblock Dynamical structure factor of electron liquid using mode-coupling
  theory,
\newblock in {\em International Centre for Theoretical Physics}, volume
  Publication IC/82/2, Trieste, 1986.

\bibitem{Kletenik11}
O.~Kletenik-Edelman, D.~R. Reichman, and E.~Rabani,
\newblock J. Chem. Phys. {\bf 134}, 044528 (2011).

\bibitem{Reichman01a}
D.~R. Reichman and E.~Rabani,
\newblock Phys. Rev. Lett. {\bf 87}, 265702 (2001).

\bibitem{Rabani02a}
E.~Rabani and D.~R. Reichman,
\newblock Phys. Rev. E {\bf 65}, 036111 (2002).

\bibitem{Reichman02a}
D.~R. Reichman and E.~Rabani,
\newblock J. Chem. Phys. {\bf 116}, 6279 (2002).

\bibitem{Rabani02b}
E.~Rabani and D.~R. Reichman,
\newblock J. Chem. Phys. {\bf 116}, 6271 (2002).

\bibitem{Rabani02c}
E.~Rabani, D.~R. Reichman, G.~Krilov, and B.~J. Berne,
\newblock Proc. Natl. Acad. Sci. USA {\bf 99}, 1129 (2002).

\bibitem{Rabani02d}
E.~Rabani and D.~R. Reichman,
\newblock Europhys. Lett. {\bf 60}, 656 (2002).

\bibitem{Rabani04}
E.~Rabani and D.~R. Reichman,
\newblock J. Chem. Phys. {\bf 120}, 1458 (2004).

\bibitem{Rabani05}
E.~Rabani, K.~Miyazaki, and D.~R. Reichman,
\newblock J. Chem. Phys. {\bf 122}, 034502 (2005).

\bibitem{Rabani05a}
E.~Rabani, G.~Krilov, D.~R. Reichman, and B.~J. Berne,
\newblock J. Chem. Phys. {\bf 123}, 184506 (2005).

\bibitem{Gotze76b}
W.~G{\"{o}}tze and M.~L{\"{u}}cke,
\newblock Phys. Rev. B {\bf 13}, 3825 (1976).

\bibitem{Jackson62}
H.~W. Jackson and E.~Feenberg,
\newblock Rev. Mod. Phys. {\bf 34}, 686 (1962).

\bibitem{Feynman56}
R.~P. Feynman and M.~Cohen,
\newblock Phys. Rev. {\bf 102}, 1189 (1956).

\bibitem{Gotze92}
W.~G{\"{o}}tze and L.~Sj{\"{o}}gren,
\newblock Rep. Progr. Phys. {\bf 55}, 241 (1992).

\bibitem{BalucaniZoppi}
U.~Balucani and M.~Zoppi,
\newblock {\em Dynamics of the Liquid State},
\newblock Oxford University Press, New York, 1994.

\bibitem{Rabani01a}
E.~Rabani and D.~R. Reichman,
\newblock J. Phys. Chem. B {\bf 105}, 6550 (2001).

\bibitem{Chandler84a}
D.~Chandler, Y.~Singh, and D.~M. Richardson,
\newblock J. Chem. Phys. {\bf 81}, 1975 (1984).

\bibitem{Chandler84b}
A.~L. {Nichols III}, D.~Chandler, Y.~Singh, and D.~M. Richardson,
\newblock J. Chem. Phys. {\bf 81}, 5109 (1984).

\bibitem{Collepardo-Guevara2008}
R.~Collepardo-Guevara, I.~R. Craig, and D.~E. Manolopoulos,
\newblock J. Chem. Phys. {\bf 128}, 144502 (2008).

\bibitem{Miller2005}
T.~F. Miller and D.~E. Manolopoulos,
\newblock J. Chem. Phys. {\bf 122}, 184503 (2005).

\bibitem{Craig2006}
I.~R. Craig and D.~E. Manolopoulos,
\newblock Chemical Physics {\bf 322}, 236  (2006).

\bibitem{Markland2008}
T.~E. Markland, S.~Habershon, and D.~E. Manolopoulos,
\newblock J. Chem. Phys. {\bf 128}, 194506 (2008).

\bibitem{Suleimanov2011}
Y.~V. Suleimanov, R.~Collepardo-Guevara, and D.~E. Manolopoulos,
\newblock J. Chem. Phys. {\bf 134}, 044131 (2011).

\bibitem{Richardson09}
J.~O. Richardson and S.~C. Althorpe,
\newblock J. Chem. Phys. {\bf 131}, 214106 (2009).

\bibitem{Kob95a}
W.~Kob and H.~C. Andersen,
\newblock Phys. Rev. E {\bf 51}, 4626 (1995).

\bibitem{Kob95b}
W.~Kob and H.~C. Andersen,
\newblock Phys. Rev. E {\bf 52}, 4134 (1995).

\bibitem{Ceriotti2010}
M.~Ceriotti, M.~Parrinello, T.~E. Markland, and D.~E. Manolopoulos,
\newblock J. Chem. Phys. {\bf 133}, 124104 (2010).

\bibitem{Parrinello1984}
M.~Parrinello and A.~Rahman,
\newblock J. Chem. Phys. {\bf 80}, 860 (1984).

\bibitem{Voth96r}
G.~A. Voth,
\newblock Adv. Chem. Phys. {\bf XCIII}, 135 (1996).

\bibitem{Voth96}
M.~Pavese and G.~A. Voth,
\newblock Chem. Phys. Lett. {\bf 249}, 231 (1996).

\bibitem{Voth04}
T.~D. Hone and G.~A. Voth,
\newblock J. Chem. Phys. {\bf 121}, 6412 (2004).

\bibitem{Leung1994}
K.~Leung and D.~Chandler,
\newblock Phys. Rev. E {\bf 49}, 2851 (1994).

\bibitem{Wipf1997}
H.~Wipf, editor,
\newblock {\em Hydrogen in Metals III: Properties and Applications},
\newblock Springer-Verlag, Berlin, 1997.

\end{thebibliography}

\end{document}